\documentclass[prd,preprintnumbers,superscriptaddress,nofootinbib,floatfix,twocolumn,notitlepage]{revtex4}
\usepackage[dvips]{graphicx}
\usepackage{dcolumn} 
\usepackage{bm}
\usepackage{epsfig,amsmath,amssymb,verbatim,mathrsfs,array,layout,textcomp,amssymb,latexsym,slashed}
\usepackage{color}
\usepackage{hyperref}
\usepackage[utf8]{inputenc}

\input epsf.tex
\newcommand{\beq}{\begin{eqnarray}}
\newcommand{\eeq}{\end{eqnarray}}

\def\ltap{\ \raise.3ex\hbox{$<$\kern-.75em\lower1ex\hbox{$\sim$}}\ }
\def\gtap{\ \raise.3ex\hbox{$>$\kern-.75em\lower1ex\hbox{$\sim$}}\ }

\def\be{\begin{equation}}
\def\ee{\end{equation}}
\def\bea{\begin{eqnarray}}
\def\eea{\end{eqnarray}}

\pdfoutput=1

\newcommand{\hc}{{\rm h.c.}}

\definecolor{newred}{rgb}{0.5,0.1,0}
\definecolor{darkgreen}{rgb}{0.0,0.7,0.2}

\begin{document} 

\title{A Dark Sector for $g_\mu-2$, $R_K$ and a Diphoton Resonance}

\author{Genevi\`eve B\'elanger}
\email{belanger@lapth.cnrs.fr}
\affiliation{LAPTh, Universit\'e Savoie Mont Blanc, CNRS B.P. 110, F-74941 Annecy-le-Vieux, France}
\author{C\'edric Delaunay}
\email{delaunay@lapth.cnrs.fr}
\affiliation{LAPTh, Universit\'e Savoie Mont Blanc, CNRS B.P. 110, F-74941 Annecy-le-Vieux, France}
\begin{flushright}
\preprint{\scriptsize LAPTH-011/16\vspace*{.1cm}}
\end{flushright}
\vskip .025in

\begin{abstract}
We revisit a set of dark sector models, motivated by  anomalies observed in $B$ decays and the muon  anomalous magnetic moment, in the light of a recently reported diphoton excess around 750$\,$GeV. Interpreting the excess as a scalar resonance associated with the symmetry breaking sector of a dark gauge group, we show that a diphoton cross section of few fb can be accomodated, together with anomalies in $R_K$ and $g_\mu-2$ within a minimal dark sector model. The resulting prominent collider signatures are in the form of wide resonant signals into top and muon pair final states below $\sim1\,$TeV. The model further predicts a dark matter candidate, yet with a significantly underabundant relic density, unless produced by an appropriate non-thermal mechanism.
\end{abstract}

\maketitle

\section{Introduction}

Nature seems malicious. While naturalness signals are lacking at the Large Hadron Collider (LHC), the ATLAS~\cite{ATLAS-dec2015} and CMS~\cite{CMS-dec2015} collaborations recently reported an excess in the diphoton invariant mass distribution around 750 GeV with a (local) significance of $3.9\sigma$ and $2.6\sigma$, respectively. Numerous interpretations of such a signal are possible, see {\it e.g.}~\cite{Franceschini:2015kwy,Falkowski:2015swt,McDermott:2015sck,Gupta:2015zzs,Ellis:2015oso}.  The simplest explanation is the production of a new resonance $X$ of spin zero or two~\cite{Landau,Yang}. 
The number of events observed in excess of the diphoton background corresponds to a signal cross section of
\beq\label{excess}
\sigma(pp\to X\to\gamma\gamma)\sim 5\,{\rm fb}\,,
\eeq
at $\sqrt{s}=13\,{\rm TeV}$. Such a large cross section requires the resonance to be accompanied by new particles~\cite{Weizmann1} in order to enhance the diphoton branching ratio (BR)  of $X$ and most likely also its production rate. A putative diphoton signal only provides rather limited decisive information regarding the associated underlying physics beyond the SM, although some interesting constraints could be derived on its couplings to SM states~\cite{Gupta:2015zzs,Weizmann1,Goertz:2015nkp,Delaunay:2016zmu,Frugiuele:2016rii}. Most notably,  nothing in current data clearly indicates whether such a resonance is playing a role in electroweak symmetry breaking, except perhaps a fortuitous resemblance with the Higgs boson discovery~\cite{ATLASdisco,CMSdisco}. 

A possible strategy towards a less ambiguous interpretation of the diphoton resonance might be to combine it with other intriguing anomalies persisting in the low-energy data. 
Here, we consider a minimal model that was initially introduced to explain a series of anomalies in observables  involving muons~\cite{Belanger:2015nma}. (See Refs.~\cite{Bauer:2015knc,Bauer:2015boy} for an alternative approach.) This includes the long-standing  $3\sigma$ deviation in $a_\mu=(g_\mu-2)/2$, the anomalous magnetic moment of the muon~\cite{gm2BNL,PDG},
\beq\label{gminus2}
\Delta a_\mu = a_\mu^{\rm exp}-a_\mu^{\rm SM}=(287\pm80)\times 10^{-11}\,,
\eeq
  as well as a recent deviation 
 in the ratio $R_K={\rm BR}(B^+\to K^+\mu^+\mu^-)/{\rm BR}(B^+\to K^+e^+e^-)$. The latter was observed about $2.6\,\sigma$   below
 the theoretically clean SM prediction of $R_K^{\rm SM}-1\sim 10^{-4}$~\cite{LHCbRK},
\beq\label{RK}
R_K^{\rm exp} =0.745^{+0.090}_{-0.074}({\rm stat})\pm0.036({\rm syst}).
\eeq
 pointing to sources of new physics in $b\to s \mu^+\mu^-$ transitions at short distances also supported by measurements of decay rate for $B^0\to K^{0(*)}\mu^+\mu^-$~\cite{Aaij:2014pli} and $B^0\to \phi \mu^+\mu^-$~\cite{Aaij:2013aln}, as well as angular distributions in $B^0\to K^{0*}\mu^+\mu^-$~\cite{Kstar}.
Several scenarios were proposed to explain the $g_\mu-2$ and $b\to s\mu\mu$ results~\cite{Belanger:2015nma,Bauer:2015knc,Gripaios:2015gra,Allanach:2015gkd}. An interesting possibility consists of introducing a massive $Z'$ vector boson which
 couples to muon pairs
 only {\it radiatively}, such that the same interaction induces the required new contributions to both $b\to s \mu\mu$  transitions and $g_{\mu}-2$~\cite{Belanger:2015nma}. 
Such a scenario calls for
 additional states, in particular new scalars and vector-like fermions, one of which could play the role of dark matter (DM). Moreover, the annihilation cross section at thermal decoupling for scalar DM is typically dominated by the very same interaction responsible for the muon-related anomalies, thus resulting in a cosmological relic close to the observed value.
Finally,  the scalar excitation in the direction of the  vacuum expectation value (VEV) generating the $Z'$ mass could be identified to the 750$\,$GeV resonance hinted at ATLAS and CMS (see also Refs.~\cite{Ko:2016lai,Ko:2016wce,Yu:2016lof}), with couplings to gluons and photons conveniently enhanced by the vector-like quarks inherent in the model. Hence, the phenomenological model proposed in Ref.~\cite{Belanger:2015nma} has all the ingredients to further explain the diphoton excess.
 
Our aim in this paper is to investigate whether this simple model  which explain two muon-related anomalies
can reproduce the large cross section required to fit the diphoton excess while satisfying all current collider constraints. 
Since the anomalies require a $Z'$ below the TeV scale, one of the most important constraint on the model comes from searches for $Z'$, notably  in the $t\bar{t}$  and $\mu^+\mu^-$ channels.
We show that  all the above collider hints of new physics  can be explained simultaneously, but only at the expense of a largely underabundant dark matter,  unless the observed relic is set through some non-thermal mechanisms.

\section{The model}
~\label{sec:model}

The model is a simple extension of the Standard Model (SM) with a new sector charged under a U(1)$_X$ gauge group, under which the SM states are neutral. We will refer to all particles with $X\neq 0$ as "dark". Besides the 
U(1)$_X$ gauge field $Z'_\mu$, the new sector consists of a vector-like quark doublets $Q,Q^c$ and a pair of vector-like lepton doublets $L,L^c$ and $\tilde L,\tilde L^c$ for each generation, as well as  two SM singlet complex scalars $\phi$ and $\chi$. The charges of the new particles under the SM and $U(1)_X$ groups are listed in Table~\ref{tab:particles}. Note that the ratio of heavy lepton to heavy quark numbers is inversely proportional to their $X$ charge ratio, thus keeping radiative contributions to the kinetic mixing of $U(1)_X$ with the SM hypercharge insensitive to unknown dynamics at the high scales~\cite{HoldomU1}. 

\begin{table}[!t]
\renewcommand{\arraystretch}{1.3}
\begin{tabular}{|c|c|ccc|c|}
\hline
  field & spin & SU(3)$_c$ & SU(2)$_L$ & U(1)$_Y$ &  U(1)$_X$ \\
\hline
$L$, $L^c$ & $1/2$& ${\bf1}$ &${\bf 2}$ & $-1/2$ & $1$\rule{0pt}{2.6ex}\\
$\tilde L$, $\tilde L^c$ & $1/2$& ${\bf1}$ &${\bf 2}$ & $-1/2$ & $1$\rule{0pt}{2.6ex}\\
$Q$, $Q^c$ & $1/2 $ & {\bf 3} & ${\bf 2}$ & $1/6 $ & $-2$\\
$\phi $ & 0&$ {\bf 1}$&${\bf 1}$ & $0$ & $2$\\
$\chi $ & 0&$ {\bf 1}$ & ${\bf 1}$
 & $0$ & $-1$  \rule[-1.2ex]{0pt}{0pt}\\
\hline
\end{tabular}
\label{tab:particles}
\caption{Dark sector states and their quantum numbers.}\vspace*{0.2cm}
\end{table}

The Lagrangian of the model is
$\mathcal{L}=\mathcal{L}_{\rm SM} + \mathcal{L}_{\rm dark} - \mathcal{L}_{\rm portal}$, 
where $\mathcal{L}_{\rm SM}$ denotes the SM part,  and
\beq
\label{eq:lagrangian}
\mathcal{L}_{\rm dark}&=& |D_\mu \phi|^2+|D_\mu \chi|^2 - V(\phi,\chi)\nonumber\\
&&+\sum_{F=Q,L,\tilde L}\bar F (i\slashed D -M_F) F\,,
\eeq
is the dark sector part (flavor indices are understood), while
\beq\label{portal}
\mathcal{L}_{\rm portal}&=& \epsilon B_{\mu\nu}X^{\mu\nu}+\lambda_{H\chi} |H|^2|\chi|^2+\lambda_{H\phi} |H|^2|\phi|^2\nonumber\\
&&+w (\bar q Q)\phi + y (\bar l L)\chi +\tilde y(\bar l\tilde L)\chi+\hc\,,
\eeq
gathers the SM interactions with the dark sector.
Here $l_i (q_i)$ are the LH SM lepton (quark) doublets and $i,j=1,2,3$ are the generation indices.
The scalar potential reads
\beq
V(\phi,\chi)&=&m_\phi^2|\phi|^2+\lambda_\phi |\phi|^4+m_\chi^2|\chi|^2+\lambda_{\chi}|\chi|^4\nonumber\\
&&+\lambda_{\phi\chi}|\phi|^2|\chi|^2+(r\phi \chi^2+\hc)\,.\label{potential}
\eeq
We assume that the potential is such that $\phi$ develops a non-zero VEV, thus breaking U(1)$_X$ and generating a $Z'$ mass of  $m_{Z'}=2\sqrt{2}g'\langle \phi\rangle$. This  requires  $m_\phi^2+\lambda_{H\phi}v^2/2<0$, with the Higgs VEV $v\simeq246\,$GeV. In order to maintain a discrete $Z_2$ symmetry in the broken phase, we further impose that $\chi$ has no VEV, leading to the conditions
$m_\chi^2+\lambda_{H\chi}v^2/2+\lambda_{\phi\chi}\langle\phi\rangle^2\pm r\langle\phi\rangle>0$.
The scalar eigenstates then read
$\phi-\langle\phi\rangle\equiv 1/\sqrt{2}(\varphi + i a)$ and
$\chi \equiv1/\sqrt{2}(\chi_0+i\chi')$.
The pseudoscalar $a$ is a spurious field associated with the longitudinal component of the massive $Z'$, while $\varphi$ is identified with the  diphoton resonance ATLAS and CMS hinted, $m_\varphi\approx 750\,$GeV. 
The scalar and pseudoscalar components of $\chi$ display a mass gap of
\beq
\delta \equiv  \frac{m_{\chi'}^2}{m_{\chi_0}^2} -1 = -4\frac{r\langle \phi\rangle}{m_{\chi_0}^2}\,.
\eeq
The $\phi$ kinetic term provides a coupling of $\varphi$ to $Z'$ pairs of $4g'm_{Z'}$, while 
$\varphi$ couplings to $\chi_0$ and $\chi'$ pairs are $m_{Z'}/(2g')[\lambda_{\phi\chi}\pm2\delta (g'm_{\chi_0})^2/m_{Z'}^2]$, respectively. 
 At tree-level, the scalar $\varphi$ also couples to the SM Higgs and quarks. Note that, by construction $\varphi$ does not couple classically to leptons, nor to SM gauge bosons.

The $X$ charge assignment of Table~\ref{tab:particles} forbids $\chi_0$ and $\chi'$ couplings to quarks.    
Although their couplings to leptons  $y,\tilde y$ could lead {\it a priori} to lepton flavor violating effects, we assume here that both are aligned with the SM charged lepton masses. Furthermore, we take for simplicity  $\tilde y=y$ and $M_{\tilde L}=M_L$.  
Motivated by the observed lepton non-universality (LNU) in semileptonic $B$ decays, we further take $y_e\ll y_\mu$, while $y_\tau$ is left unspecified. This captures the relevant ingredients required to address the $R_K$ anomaly.

The $Z'$ couples to SM fermions only through U(1)$_X$ breaking effects. $Z'$ couplings to SM quark doublets arise at tree-level through mass mixing with $Q$, while there is no mixing with SM singlets. This results in {\it partially dark} left-handed (LH) SM quarks with, schematically, $q_{SM}\equiv \cos\theta\, q + \sin\theta\, Q_L$, the angle $\theta$ measuring their ``degree of darkness''. Explaining $b\rightarrow s\mu\mu$ anomalies requires flavour mixing in the quark sector, in particular a flavour changing $Z'$ coupling between $b_L$ and $s_L$ quarks. This is realized through an appropriate flavour structure in the dark quark spectrum $M_Q$ and the portal coupling $w$. A sufficient set of conditions is that {\it i)} the unitary matrix $V_Q$, which parameterizes the misalignment between $w^\dagger w$ and $M_Q^\dagger M_Q$, is not proportional to the identity matrix and {\it ii)} $M_Q$ is not degenerate. For simplicity, we limit ourselves to mixings among the second and third quark generations and neglect possible CP-violating phases. Then, as shown in Appendix~\ref{quarkmixing}, $V_Q$ is simply an orthogonal matrix of angle $\theta_Q$ and the tree-level $Z'$ couplings to LH quark currents are (in units of $g'$)
\beq
g_{\bar s s,\bar bb}^{Z'}&=&2\left(\cos^2\theta_Q\sin^2\theta_{s,b}+\sin^2\theta_Q\sin^2\theta_{b,s}\right)\,,\label{eqg1}\\
g_{\bar b s}^{Z'}&=&\sin(2\theta_Q)(\sin^2\theta_s-\sin^2\theta_b)\label{eqg2}\,.
\eeq
$\sin\theta_{b,s}\simeq w\langle\phi\rangle/m_{Q_{3,2}}$ are the sines of the partial darkness angle of LH bottom and strange quarks, while $m_{Q_{3,2}}\simeq(M_{Q_{3,2}}^2+w^2\langle\phi\rangle^2)^{1/2}$ are the heavy quark eigenmasses. Note that, $\sin\theta_s/\sin\theta_b=m_{Q_3}/m_{Q_2}$ as a result of the assumed $w$ degeneracy. Moreover, since the mass mixing preserves SU(2)$_L$, the same coupling structure applies to the up-quark sector. 

In constrast, the $Z'$ does not couple to SM leptons at tree-level due to the absence of mixing with heavy leptons. For $\delta\neq0$, such a coupling is induced at one-loop level with $g_{\bar \mu\mu}^{Z'}=y^2/16\pi^2F(\tau,\delta)$ where $F$ is a loop function defined in~\cite{Belanger:2015nma},  $\tau\equiv M_L^2/m_\chi^2$. As long as $M_{L}\lesssim 800\,$GeV and $\delta\gg 1$, this allows to simultaneously explain the $b\to s\mu\mu$ and $g_\mu-2$ anomalies~\cite{Belanger:2015nma}.

Finally, consider the $\varphi$ interactions with quarks. Since the right-handed (RH) SM quarks do not mix with $Q$, there is no $\varphi \bar{q}q$ couplings in the mass basis, but only $\varphi \bar{q}Q$ and $\varphi \bar{Q}Q$ couplings. For the flavour structure assumed above, those are 
\beq
g^{\varphi }_{\bar s Q_2}= g^{\varphi }_{\bar b Q_3} \frac{\cos\theta_s}{\cos\theta_b} =   g'\sin(2\theta_s)\frac{m_{Q_2}}{m_{Z'}}
\eeq
and
\beq 
g^{\varphi}_{\bar {Q}_{2} Q_{2}} = g^{\varphi}_{\bar {Q}_{3} Q_{3}}\frac{\sin\theta_s}{\sin\theta_b} = 2g'\sin^2\theta_{s}\frac{ m_{Q_2}}{m_{Z'}}\,,
\eeq
 respectively, with similar expressions for charm and top quarks.
Note  the absence of  flavor changing $\varphi$ couplings, which is a mere consequence of the assumed flavour universality of $w$. The $\varphi \bar QQ$ couplings will play a crucial role in producing $\varphi$ through gluon fusion with a large rate, as well as enhancing its branching ratio in diphotons, in particular for $m_{Q_{2,3}}\sim m_\varphi$.

\section{Addressing collider anomalies}\label{required}

Consider first the sizable resonant cross section of $\varphi$  in the diphoton channel in Eq.~\eqref{excess}.
Since $\varphi$ does not couple at tree-level to SM quark pairs, nor to vector boson pairs,  its production at the LHC is dominated by gluon fusion through a loop of both isospin components of the vector-like quarks $Q_{2,3}$. Using the generic expressions for loop-induced scalar couplings to a pair  of gluons, see {\it e.g.} Ref.~\cite{anatomy}, and MSTW parton distribution functions~\cite{MSTW} at leading order, we find 
\beq\label{phiprod}
\sigma_{13}(pp\to\varphi) &\approx &0.35\,{\rm pb}\times k_{\rm QCD}\left(\frac{650\,{\rm GeV}}{m_{Z'}/g'}\right)^2\nonumber\\
&&\times\sin^4\theta_b\left|1+I\left(\frac{\sin\theta_s}{\sin\theta_b}\right)\right|^2\,, 
\eeq
for $m_{Q_3}= m_{Q_2}(\sin\theta_s/\sin\theta_b)=800\,$GeV. 
 $I(x)$, with $x\equiv \sin\theta_s/\sin\theta_b$, is the $Q_2$-to-$Q_3$ loop amplitude ratio, satisfying $I(1)=1$ and $I(x)\simeq x^3$ for $x\ll1$,  and the multiplicative factor $k_{\rm QCD}$ captures unspecified higher-order QCD corrections.
Hence, the branching ratio $\varphi\rightarrow \gamma\gamma$ must be at least ${\cal O}(10^{-3}$) in order to reach the high signal rate of Eq.~\ref{excess}. In the minimal setup considered in this letter, this is only achievable by {\it i)} boosting the $\gamma\gamma$ partial width with loops of  $Q_{2,3}$ and, most importantly, {\it ii)} suppressing all tree-level two-body decays, in particular $\varphi\to Z'Z'$, $\varphi\to \chi\chi$ and $\varphi\to qQ$, as well as potentially sizable  three-body decays into $Z'\bar q q$ ($q=s,c,b,t$). Avoiding overly-large widths in tree-level two-body decays requires $m_{Z'} > m_\varphi/2$, $m_{\chi} > m_\varphi/2$ and $m_Q> m_\varphi$. 
Three-body decays into $Z'\bar q q$ arise through either one off-shell $Z'$ ( $\varphi\rightarrow Z'Z'^*\rightarrow  Z' q\bar{q}$)  or one off-shell $Q$ ($\varphi\rightarrow \bar qQ^*\rightarrow \bar{q} qZ'$). Both can give important contributions to the total width, thus imposing a further constraint on the $Z'$ mass. While the later depends on the couplings involved it roughly requires $m_{Z'} \gtrsim 650 {\rm GeV}$. 
In this case, $\varphi$ is found to decay dominantly into gluons, with $\Gamma_\varphi\simeq\Gamma(\varphi\to gg)\sim\mathcal{O}(1)\,$GeV, while
the diphoton partial width is found to be
\beq
\Gamma(\varphi\to\gamma\gamma)&\approx& 0.77\,{\rm MeV}\times \left(\frac{650\,{\rm GeV}}{m_{Z'}/g'}\right)^2\nonumber\\
&&\times\sin^4\theta_b\left|1+I\left(\frac{\sin\theta_s}{\sin\theta_b}\right)\right|^2\,,
\eeq
corresponding to BR$(\varphi\to \gamma\gamma)\sim\mathcal{O}(10^{-3})$. 
Combining the above with Eq.~\eqref{phiprod} gives a diphoton signal in the ballpark of Eq.~\eqref{excess} provided $\sin\theta_{b,s}\approx1$.

The second request is to explain both the $R_K$ and $g_\mu-2$ anomalies. $R_K$ is addressed by a $Z'$ mediated contribution to the $(\bar b s)_{V-A}(\bar \mu\mu)_{V-A}$ local operator at the bottom mass scale. The flavor changing $Z'$ coupling sourcing the new $b\to s\mu\mu$ amplitude is strongly constrained by $\bar B_s-B_s$ oscillations. Restricting  the $Z'$ contribution to the mass difference of neutral $B_s$ mesons to $10\%$ of the SM contribution gives $g^{Z'}_{\bar b s}\lesssim5.2\times 10^{-3}(m_{Z'}/650\,$GeV$)/g'$~\cite{Altmannshofer-Straub,Buras-DeFazio-Girrbach}. As a result, the central value of $R_K$ in Eq.~\eqref{RK} requires a rather large $Z'$-to-muon coupling of $g^{Z'}_{\bar \mu \mu}\gtrsim6.3\times 10^{-2}(m_{Z'}/650\,$GeV$)/g'$~\cite{Belanger:2015nma}. This corresponds to a relatively strong leptonic coupling  
\beq\label{ylowerbound}
y\gtrsim 7.6\left(\frac{m_{Z'}/g'}{650\,{\rm GeV}}\right)^{1/2}\left[\frac{0.17}{F(\tau,\delta)}\right]^{1/2}\,.
\eeq
Through a loop of $\chi$ and vector-like leptons, the same interaction induces a $g_\mu-2$ contribution of $\Delta a_\mu= y^2m_\mu^2/m_{\chi_0}^2 G(\tau,\delta)$, where $G(\tau,\delta)$ is a loop function defined in~\cite{Belanger:2015nma}. Saturating the lower bound in Eq.~\eqref{ylowerbound}, the discrepancy in Eq.~\eqref{gminus2} is accomodated for $m_{\chi_0}\approx 400\,$GeV, which is consistent with $m_{\chi_0}>m_\varphi/2$ necessary to maintain a large ${\rm BR}(\varphi\to \gamma\gamma)$.

In order to assess the robustness of the above discussion, we varied the input parameters of the model within
$0.5< g'< 4$, $1<y<4\pi$, $550< m_{Z'}< 950\,$GeV,  $750< m_{Q_{2,3}} < 1800\,$GeV,
 $375< m_{\chi_0}< 800\,$GeV,  $1< M_{L}/m_{\chi_0}< 3$, $0<\delta < 20$, $0<\sin\theta_{s,b}<1$ and 
$|\sin\theta_{Q}| < 1$. The parameters of the Higgs potential do not enter the observables directly and $\lambda_{\phi\chi}$ does not play an important role here, thus we simply set them to zero. We will comment latter on their possible impact when discussing DM observables.
For each point in parameter space, we computed the diphoton cross section of the $\varphi$ scalar, $g_\mu-2$ and $R_K$. Cross sections and DM observables were computed using {\tt micrOMEGAs 4}~\cite{micrOMEGAs41} and {\tt CalcHEP 3.4}~\cite{calchep}. Our main result is displayed in Fig.~\ref{fig:gamgam_tt} which shows that the observed diphoton rate can be accomodated, together with $b\to s\mu\mu$ and $g_\mu-2$ anomalies. However, for $\sigma_{13}(pp\to S\to \gamma\gamma)\gtrsim4{\rm fb}$, there is a tension with concomitant resonance searches in top and muon pairs as discussed next.

\section{Collider constraints and signatures}\label{collider}

The $Z'$ necessarily have sizable couplings to second and third generation quark pairs. Indeed, those are related to the $\varphi$-to-vector quark couplings which needs to be $\mathcal{O}(1)$ to sustain a large diphoton signal. As a result, the model potentially faces several constraints from various resonance searches at the LHC. Consider for illustration the case of a $650\,$GeV $Z'$. With MSTW PDFs~\cite{MSTW}, the leading order production cross section at the $8\,$TeV LHC is
\beq
\sigma_{8}(pp\to Z')\approx 56\,{\rm pb}\times g^{\prime 2}\left(\sin^4\theta_s+0.13\sin^4\theta_b\right),
\eeq
where the first (second) term in parenthesis represents the contribution from $\bar cc+\bar ss$ ($\bar bb$) annihilation. The cross section  is  a factor of $\sim5$ larger at $13\,$TeV. With a typical $\mathcal{O}(1)$ BR into jet pairs, this signal is generically larger than that probed in dijet resonance searches in this mass range by an order of magnitude~\cite{Aad:2014aqa}. However, note that for $\sin\theta_b\simeq\sin\theta_s\approx 1$ the $Z'$ is unavoidably wide. With typically $\Gamma_{Z'}/ m_{Z'} \gtrsim 30\%$, the above searches, which strongly rely on bump-hunting techniques, are expected to be less sensitive to such broad $Z'$ states. 

\begin{figure}[t]
\includegraphics[width=0.47\textwidth]{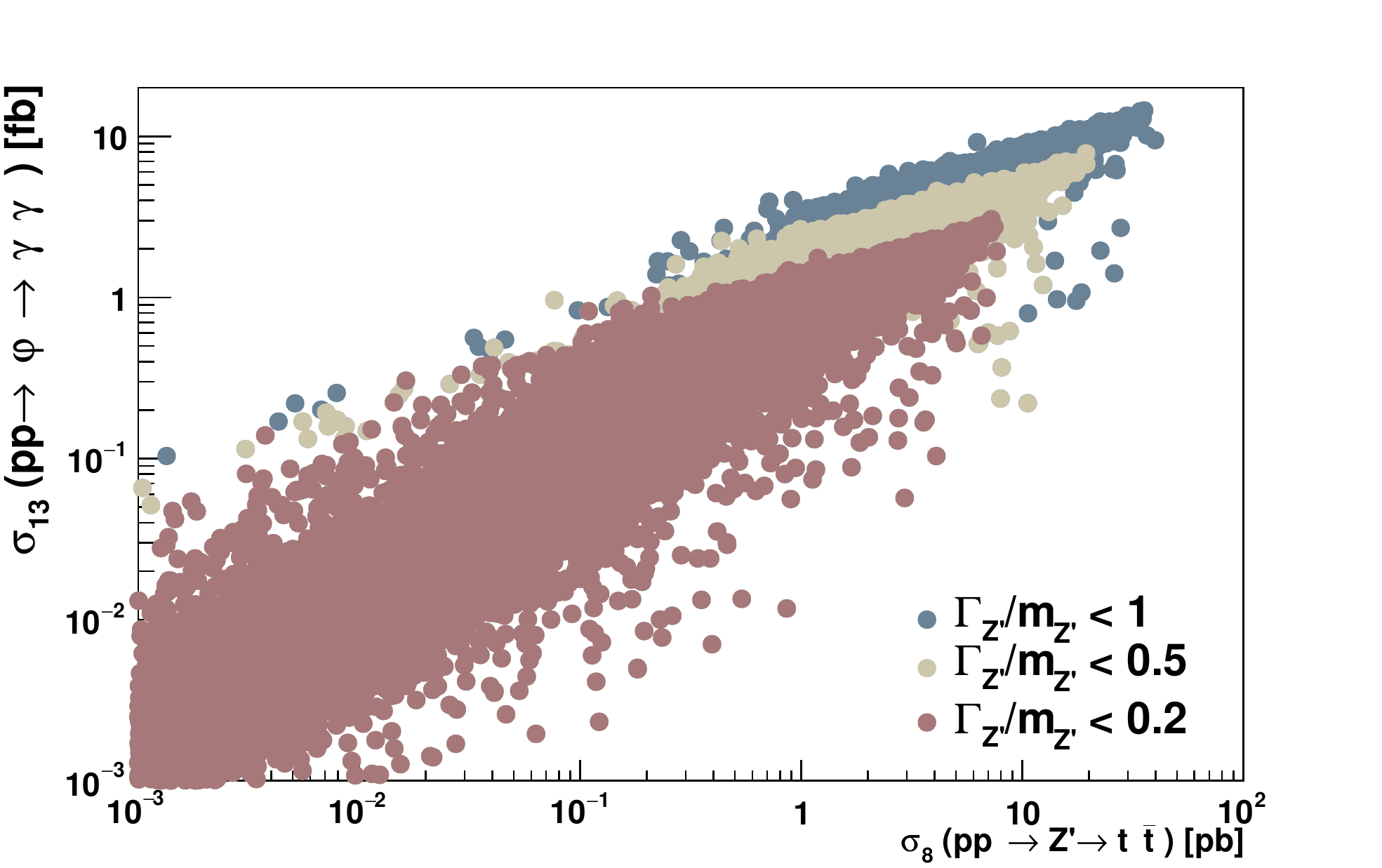}
\caption{Correlation between the 750$\,$GeV diphoton signal strength and the top pair production cross section from resonant $Z'$, for different $Z'$ width regime.}
\label{fig:gamgam_tt}
\end{figure}

\begin{figure}[t]
\includegraphics[width=0.47\textwidth]{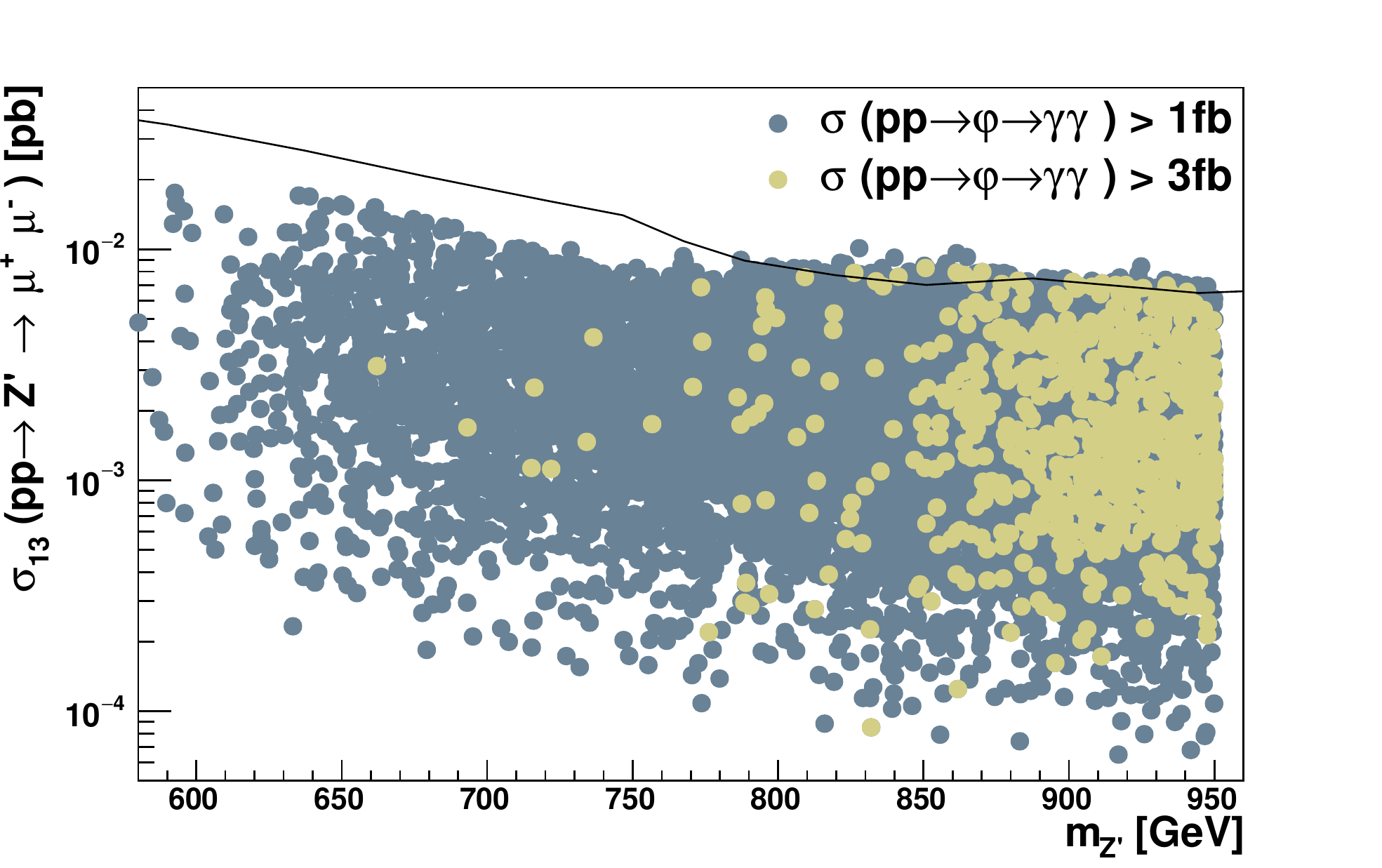}
\caption{Dimuon production cross section  through a resonant $Z'$ as a function of the $Z'$ mass  at the 13$\,$TeV LHC. All points comply with $Z'$ searches in the dimuon and $t\bar{t}$ channels at $8\,$TeV, as well as with a diphoton signal strength above $1\,$fb (blue) or $3\,$fb (yellow). The black line is the  $95\%$~CL exclusion bound from current LHC data.}\label{fig:Zpmumu}
\end{figure}

  On the other hand, a significant constraint arises from $t\bar{t}$ resonance searches, which set an upper limit of $\sim2\,$pb for $650\,$GeV vector resonances as wide as $40\%$~\cite{Aad:2015fna}. 
This constraint is easily avoided by taking either $\sin\theta_b\ll1$, thus suppressing the $Z'\to \bar tt$ decay, or $\sin\theta_s\ll 1$, thus reducing the overall $Z'$ production cross section. Consequently, $\varphi$ couples mostly to second or third generation quarks, but not both. Either way, this results in smaller loop-induced $\varphi$ couplings to gluons and photons which limits the diphoton signal to $\simeq4\,$fb, as shown in Fig.~\ref{fig:gamgam_tt}. A cleaner $Z'$ signature is in the dimuon channel. However, the associated partial width is loop-induced with typically ${\rm BR}(Z'\to\mu^+\mu^-)\sim\mathcal{O}(10^{-3})$ or less. This results in $8\,$TeV cross sections as large as $\mathcal{O}(0.1)\,$pb, which is in an order of magnitude tension with current data~\cite{Aad:2014cka}.
Again, this can be avoided for $\sin\theta_s\ll 1$. 
Note that, since the partial $Z'$ width in muon pairs is independent of $\sin\theta_{b,s}$, there is no correlation between the dimuon channel and the diphoton cross section. Imposing a diphoton signal strength above $1\,$fb, as well as the $\bar tt$ and $\mu^+\mu^-$ upper limits from the 8$\,$TeV data, the 13$\,$TeV dimuon cross section is still allowed by current data~\cite{ATLAS-CONF-2015-070}, see Fig.~\ref{fig:Zpmumu}. 

The model also predicts collider signatures involving the new vector-like quarks and leptons. EW production of heavy lepton pairs leads to $\mu^+\mu^-$ with missing energy in the final state, which is constrained by LHC data~\cite{Aad:2014vma} unless $M_L>450$GeV and/or $M_L-m_{\chi_0}\lesssim60\,$GeV~\cite{Belanger:2015nma}. 
Heavy quarks are mostly pair produced through QCD interactions. When decays into an EW state
($W,Z,h$) and a jet are dominant, strong limits of  $\sim600-900\,$GeV 
typically apply~\cite{Sahinsoy:2057733,Khachatryan:2015oba,Khachatryan:2015gza}. However, in most of the favoured parameter space, heavy quarks escape these limits since they preferably decay into $Q\rightarrow Z' q$ or
 $Q\rightarrow\varphi q$, leading to distinctive final states with leptons and/or multiple jet.

Finally, the model predicts further signatures of the 750$\,$GeV scalar. In addition to the dominant decay mode  into gluon pairs,  $\varphi$ can decay into
  $Z'q\bar{q}$, however this branching ratio is at most 10\% for points with a large diphoton cross section and it drops rapidly with $m_{Z'}$. Resonant signals in $Z\gamma$, $ZZ$, $WW$ with cross sections of the same order as the diphoton rate are also expected.
 Moreover, there is a possible monojet signal from the $s$-channel exchange of an off-shell $\varphi$, yet with a cross section which is at least one order of magnitude below current LHC sensitivity~\cite{Barducci:2015gtd,Aad:2014nra}. By SU(2)$_L$ invariance, the model predicts $t\to c$ decays through loops of $Z'$ and quarks. Besides the loop factor suppression, these FCNC signals suffer a strong GIM-like suppression. For instance, the $Zt_Lc_L$ coupling is typically $\mathcal{O}[V_{tb}^*V_{cs}g^{Z'}_{\bar b s}g^{Z'}_{\bar b b}m_t^2/(m_{Z'}/g')^2/16\pi^2]$. Saturating the $B_s-\bar{B_s}$ mixing bound on $g_{\bar b s}^{Z'}$ gives BR$(t\to cZ)\sim10^{-11}$, for $m_{Z'}/g'=650\,$GeV, which is not observable at the LHC~\cite{Chatrchyan:2013nwa,Liss:1564937}.

\section{Dark Matter Implications}\label{DM}

\begin{figure}[t]
\includegraphics[width=0.47\textwidth]{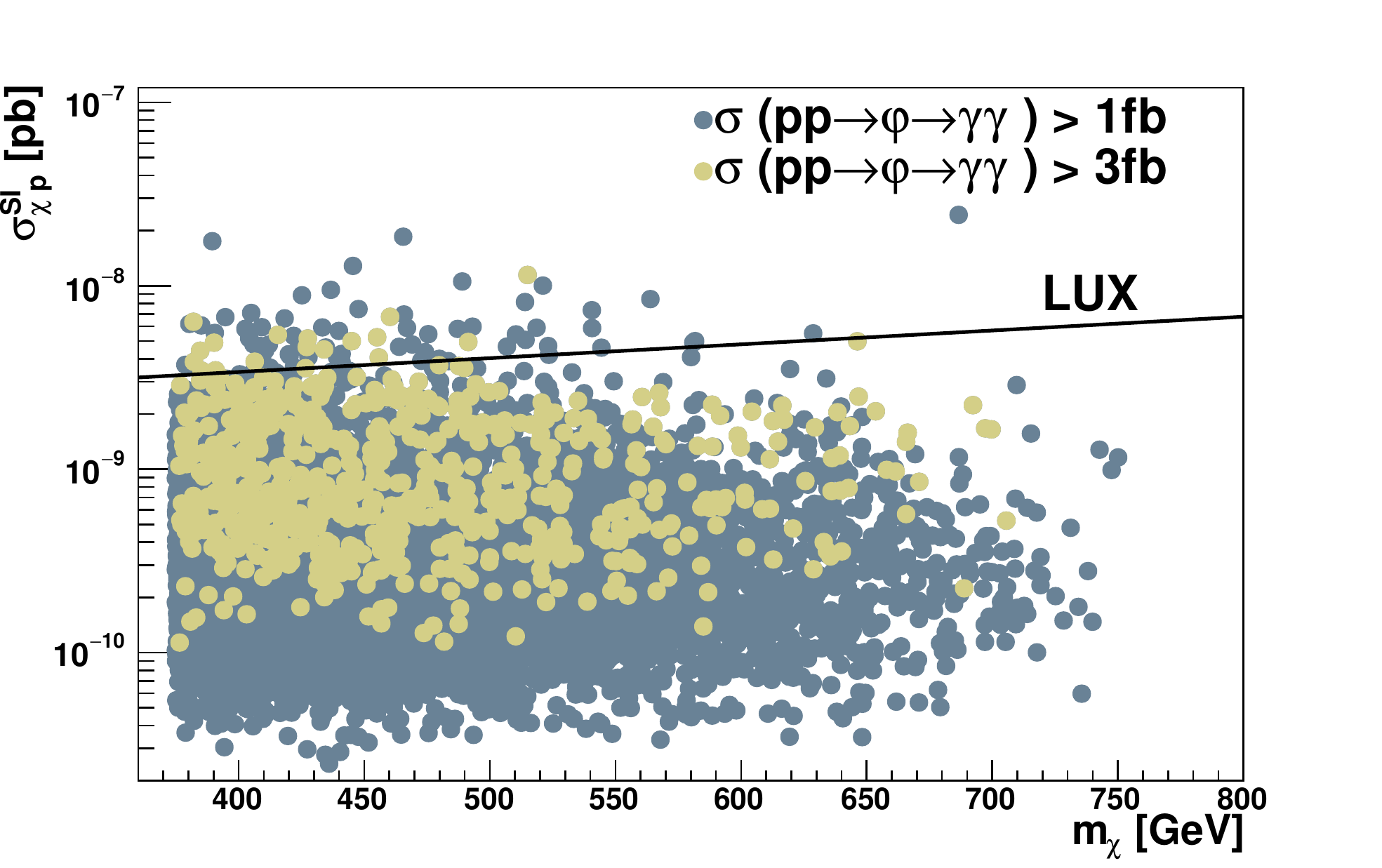}
\caption{Spin-independent cross section on protons as function of the DM mass.
The black line is the  $90\%$~CL exclusion from LUX.}\label{fig:si}
\end{figure}

The lightest dark sector state is typically the SM singlet scalar $\chi_0$. In this case, the DM annihilation in the early Universe proceeds dominantly into muon and muon-neutrino pairs through $t$-channel exchange of vector-like leptons. Under the $g_\mu-2$ and $R_K$ requests, DM was found to be significantly underdense unless $m_{\chi_0}\simeq 100\,$GeV~\cite{Belanger:2015nma}. 
In constrast, the further requirement of a large diphoton cross section implies that $m_{\chi_0}>m_\varphi/2\approx375\,$GeV. As a result, $\chi_0$ can only be a viable DM candidate if produced by an adequate non-thermal mechanism~\cite{Moroi:1999zb,Gelmini:2006pw,Baer:2014eja}\footnote{An alternate possibility is that the observed value of the relic density is mostly in the form of an another DM state. In this case, the collider anomalies under consideration are largely unrelated to DM phenomenology.}.
Assuming such a mechanism exists, direct DM detection could still probe the model. The expected cross section strongly depends on the parameters of the scalar potential in Eq.~\eqref{potential}. In particular $\lambda_{h\chi}$ controls the SM Higgs exchange contribution between DM and nuclei, while $r\propto \delta$ and $\lambda_{\phi\chi}$ set the 750$\,$GeV scalar contribution. 
Taking into account only the $\varphi$ exchange contribution proportional to $r$ and restricting to diphoton signals above $1\,$fb, we find 
spin independent cross sections as large as $\sim10^{-8}\,$pb for $m_{\chi_0}\approx400\,$GeV, which is in tension with limits from the LUX experiment~\cite{Akerib:2013tjd}, see Fig.~\ref{fig:si}.
Furthermore, barring cancellations with other contributions to the spin independent cross section, future direct detection experiments will probe the bulk of the parameter space. 

Another possiblity is that the DM is identified with the neutral vector-like lepton ($\tau<1$). However, in this case, the sizable DM interactions with the SM $Z$ and the $Z'$ states yields an overly suppressed relic density as well as spin-independent scattering cross sections on nuclei that exceed current bounds by several orders of magnitude~\cite{Akerib:2013tjd}. This situation is akin to the supersymmetric partner of SM neutrinos~\cite{Falk:1994es}.

\section{Conclusions}\label{conclusion}

We revisited a class of dark sector models originally proposed to simultaneously explain the reported anomalies in $g_\mu-2$ and $b\to s\mu^+\mu^-$ decays~\cite{Belanger:2015nma}. In this letter, we observe that these models can further accomodate the large diphoton excess reveiled by recent LHC data. The diphoton signal arises from a scalar resonance associated to the dark U(1)$_X$ breaking sector. While similar interpretations of the $750\,$GeV diphoton excess at the LHC exist in the literature, we show that combining the above hints of physics beyond the Standard Model significantly limits the viable region of parameter space in such models. Under the requirement of explaining the muon-related anomalies, we find that either the diphoton signal is negligible or the relic density of the dark matter is orders of magnitude smaller than the level suggested by CMB observations. This illustrates how addressing several anomalies in a coherent framework could yield valuable informations about the underlying physics at play. 

\section*{Acknowledgements}
We wish to thank Tetiana Hryn'Ova-Berger and Alexander Pukhov for discussions. This work is supported by the ``Investissements d'avenir, Labex ENIGMASS'', by the French ANR,  Project DMAstro-LHC, ANR-12-BS05-006 and by  the Research Executive Agency (REA) of the European Union un
der the Grant Agreement PITN-GA-2012-316704 (``HiggsTools"). \\

\appendix 

\section{Quark Flavor mixing}\label{quarkmixing}
 
We inspect here the flavor structure of the quark sector in greater details. Besides the SM Yukawa couplings $Y_{u,d}$, the  flavor parameters in the quark sector are the vector-like mass $M_Q$ and the portal interaction $w$, 
which transform, respectively, as $({\bf 1}, {\bf 3},{\bf \bar 3})$ and $({\bf 3}, {\bf 1},{\bf \bar 3})$ under U(3)$_q\times $U(3)$_{Q_L}\times$U(3)$_{Q_R}$ global flavor symmetries. Both $M_Q$ and $w\langle \phi\rangle$ are expected to be significantly larger than SM quark masses, so the latter can be neglected in first approximation. 
Then, without loss of generality, the relevant Lagrangian operators read
\beq
-\mathcal{L}\supset 
\bar Q^i_L M_Q^{ii} Q^i_R + \bar q^{\,i} w_{ii}(V_Q^\dagger)^{ij} Q_R^j\phi+\hc\,, 
\eeq
where $V_Q$ is a generic $3\times 3$ unitary matrix. The $R_K$ anomaly requires a new source of $b\to s$ transition, mediated here by a $Z'$ FCNC coupling. A first condition is $M_Q^{ii}$ non-degeneracy. Indeed, if $M_Q$ were universal, $V_Q$ could be rotated away by a $U(3)_{Q_L+Q_R}$ transformation. A second condition is to stay away from the alignment limit where $V_Q$ is the identity matrix. 
For the sake of simplicity, we further assume universal $w_{ii}$ and reality of $V_Q$, which is sufficient to obtain the required phenomenology. 

The operator $\phi (\bar q Q)$ induces mass mixing after U(1)$_X$ breaking. The quark mass matrix is then diagonalized by
\beq\label{LHrot}
\left(\begin{array}{c} Q_L^i \\ (V_Qq)^i \end{array}\right)\to \left(\begin{array}{cc} c_i & -s_i \\s_i & c_i\end{array}\right) \left(\begin{array}{c}  Q_L^i \\  q^i \end{array}\right)\,,
\eeq
where \beq
c_i \equiv \frac{M_Q^{ii}}{m_Q^i}\,,\quad s_i \equiv\frac{w\langle\phi\rangle}{m_Q^i}\,,
\eeq
 with $m_Q^i\equiv\sqrt{(M_Q^{ii})^2+w^2\langle\phi\rangle^2}$, denote the cosine and sine of the angle $\theta_i$ controlling the partial darkness of the SM quarks $q^i$.
 Before U(1)$_X$ breaking, the $Z'$ coupling to LH quarks reads 
\beq
g'Z'_\mu \left[\bar \psi_Q^i \gamma^\mu\left(\frac{1-\gamma_5}{2}\right) G_L^{ii} \psi_Q^i\right]\,,
\eeq
with $\psi_Q^i = (q^i,Q_L^i)^T$,
\beq
G_L^{ii} = \left(\begin{array}{cc} 0 & 0 \\ 0 & X_Q \end{array}\right)\,,
\eeq
and $X_Q=-2$. Since $G_L^{ii}$ is not proportional to the identity matrix, the $Z'$ couplings to LH SM quarks become flavour-changing in the mass basis defined by Eq.~\eqref{LHrot}.
Restricting to the second and third generations ($i=b,s$), 
\beq
V_Q=\left(\begin{array}{cc} \cos\theta_Q & \sin\theta_Q \\ -\sin\theta_Q & \cos\theta_Q \end{array}\right)\,,
\eeq
and the $Z'$ couplings to LH SM quarks are simply 
$g' Z'_\mu \bar{q} \gamma^\mu [(1-\gamma_5)/2]G_q^{Z'}q$,
\beq
 G_q^{Z'} = \left(\begin{array}{cc} g_{\bar ss}^{Z'} & g_{\bar bs}^{Z'} \\ g_{\bar bs}^{Z'} & g_{\bar bb}^{Z'} \end{array}\right)\,,
\eeq 
where the entries are given in Eqs.~\eqref{eqg1}-\eqref{eqg2}.
Note that, the flavour-changing $Z'$ coupling to LH bottom and strange quarks will be typically much smaller than its flavour diagonal counterparts if either $s_s\simeq s_b$ or $\theta_Q\simeq0$, corresponding to the degeneracy and alignment limits, respectively.

\bibliographystyle{apsrev}
\bibliography{X750}

\end{document}